\documentclass{PoS}

\def\eq#1{Eq.~(\ref{#1})}

\newcommand{\ord}{{\cal O}}
\newcommand{\RE}{{\rm Re}}

\def\beq{\begin{equation}}
\def\eeq{\end{equation}}
\def\beqa{\begin{eqnarray}}
\def\eeqa{\end{eqnarray}}
\def\ifm{\ifmmode}

\def \e {\epsilon}

\def \as {\relax\ifmmode\alpha_s\else{$\alpha_s${ }}\fi}

\def \al #1 {\frac {\as({#1})}{\pi} }
\def \ds #1 {\ooalign{$\hfil/\hfil$\crcr$#1$}}

\def\eps{\varepsilon}
\def\eq#1{Eq.~(\ref{#1})}

\title{Progress on the infrared structure of multi-particle gauge theory amplitudes}

\ShortTitle{Infrared Structure}

\author{\speaker{Lorenzo Magnea}\\
        University of Torino and INFN, Sezione di Torino\\
        E-mail: \email{lorenzo.magnea@unito.it}}

\abstract{I will review some of the recent intense activity concerning infrared and collinear 
divergences in gauge theory amplitudes. The central quantity in these studies is the 
multi-particle soft anomalous dimension matrix, which is completely known at two loops
for both massless and massive particles, and whose properties are currently being studied
at three-loops and beyond. I will describe how, in the massless case, the simple dipole-like
structure of the anomalous dimension up to two loops can be exploited in the high-energy limit 
to study effects that go beyond the standard form of Regge factorization. Furthermore, I will 
briefly review some of the techniques that have recently been developed to compute the soft 
anomalous dimension at high orders in perturbation theory, and I will give some examples 
of applications, including a result valid to all orders in perturbation theory for a specific class 
of diagrams.} 

\FullConference{ Loops and Legs in Quantum Field Theory - LL 2014,\\
		 27 April - 2 May 2014 \\
		 Weimar, Germany }

\begin{document}

\section{Introduction}
\label{intro}

The interest in the structure of infrared divergences of non-abelian gauge theory 
amplitudes has revived in recent years, due to several reasons, both phenomenological 
and theoretical. 

A practical concern is the fact that infrared singularities must be understood in great 
detail in order to construct efficient subtraction algorithms, which are needed in order 
to compute infrared-safe quantities at non-trivial orders in perturbation theory. The 
need to provide precise predictions  for complex processes such as jet production at 
LHC has driven a vast effort to push these calculations beyond NLO~\cite{Currie:2013dwa}, 
focusing interest on the infrared structure of scattering amplitudes at high orders. 

Another issue of great relevance for phenomenology is the need to resum large 
logarithmic corrections, which jeopardize the reliability of the perturbative expansion 
for many cross sections of physical interest. The technology of resummation is based 
upon the knowledge of soft and collinear singularities~\cite{Sterman:1995fz,Magnea:2008ga}: 
indeed, while actual divergences cancel in physical cross sections, they leave behind logarithmic enhancements, which can be computed to all orders in perturbation theory with increasing 
accuracy, provided the structure of singularities is sufficiently well understood. Resummations 
are well developed for cross sections involving only two colored particles at tree level, such 
as electroweak vector boson production, DIS, or Higgs production~\cite{Gardi:2007ma}. 
The needs of LHC have however generated interest in the possibility to provide resummed 
predictions for more complicated multi-particle processes. These predictions in 
turn require a deeper understanding of infrared divergences of multi-particle amplitudes. 

Finally, it must be noted that the study of long-distance singularities is of great interest 
also from a purely theoretical point of view. One aspect of this interest is the role played 
by singularities in the scattering amplitudes of $N = 4$ Super Yang-Mills (SYM)
theory, which have been the focus of massive investigations and impressive progress 
in the past several years~\cite{Elvang:2013cua}. In this case the quantum (super) conformal 
invariance of the theory makes the very idea of an $S$-matrix rather precarious: indeed, 
scattering amplitudes in $N = 4$ SYM are, to a large extent, defined by their infrared 
regularization, and infrared divergences in fact determine the entire structure of the 
perturbative expansion, at least up to the conformally invariant  `remainder functions' 
which start arising at two-loops and with at least six colored particles~\cite{DelDuca:2009au,
Goncharov:2010jf}. The all-order knowledge of infrared divergences for planar amplitudes 
has been instrumental to provide ideas and check results in this field~\cite{Bern:2005iz}, 
and it is to be hoped that the same will happen in the non-planar case as well.

More generally, infrared singularities provide a window into the all-order behavior of 
perturbative gauge theories: they are sufficiently simple and universal to be computable
with great accuracy, yet they retain highly nontrivial information about dynamics. 
Multi-particle amplitudes are especially interesting in this regard, since they display,
at the level of the anomalous dimension, highly non-trivial correlations between color
and kinematic degrees of freedom, which must ultimately be reflected (in QCD) by 
the patterns of parton evolution which prepare colored particles for hadronization.

In what follows, I will focus mostly on progress which has been achieved in the 
last couple of years on two fronts: on the one hand, the application of our understanding
of infrared structure to the high-energy limit, where one can control contributions
going beyond the simplest form of Reggeization~\cite{DelDuca:2011ae,Bret:2011xm}, 
making predictions at two and three loops~\cite{DelDuca:2013ara}; on the other hand, 
the development and application of computational tools for the calculation of the soft 
anomalous dimension at high orders~\cite{Gardi:2013saa,Falcioni:2014pka}, and in 
particular at three loops. This calculation, which would have looked unfeasible just a 
few years ago, appears now within reach.

\section{On the structure of infrared divergences}
\label{struct}

The distinguishing character of soft and collinear radiation is universality. Divergences
arise from emissions that are unsuppressed at large distances and long times, and thus
happen on energy scales which are drastically different from those associated with 
hard processes. In other words, infrared radiation does not possess the resolving
power to probe the hard scattering: it happens `much later' or `much earlier', so that
quantum-mechanical interference with the short-distance process is suppressed by powers
of the hard scale. While this picture is rather intuitive and physically appealing,
the structure of perturbation theory makes proving it very difficult~\cite{Sterman:1995fz,
Dixon:2008gr,Feige:2014wja}. When the dust settles, one is left with a simple factorization 
formula, which applies to all gauge theory amplitudes, and is exact for infrared divergent 
contributions. One writes~\cite{Becher:2009cu,Gardi:2009zv}
\beq 
  {\cal M} \left(\frac{p_i}{\mu}, \as \right) \, = \, {\cal Z} \left(\frac{p_i}{\mu}, \as \right)
  {\cal H} \left(\frac{p_i}{\mu}, \as \right) \, ,
\label{IRfact}
\eeq
where the scattering amplitude ${\cal M}$, involving $n$ colored particles with momenta 
$p_i$, is a vector in color space, while ${\cal Z}$ is a universal operator encoding 
infrared divergences, and acting upon the finite vector of ``matching coefficients'' 
${\cal H}$. The infrared operator ${\cal Z}$ obeys a renormalization group equation
which, in dimensional regularization~\cite{Magnea:1990zb} and for $d = 4 - 2 \epsilon 
> 4$, has a very simple solution, expressed in terms of an anomalous dimension 
matrix $\Gamma$ as
\beq
  {\cal Z} \left(\frac{p_i}{\mu}, \as \right) \, = \,  
  {\cal P} \exp \left[ \frac{1}{2} \int_0^{\mu^2} \frac{d \lambda^2}{\lambda^2} \, \,
  \Gamma \left(\frac{p_i}{\lambda}, \alpha_s(\lambda^2) \right) \right] \, .
\label{RGsol}
\eeq
Quite naturally, the matrix $\Gamma$ is the focus of all investigations in this field: 
its knowledge completely solves the perturbative infrared problem for all gauge 
theory scattering amplitudes. Currently, $\Gamma$ is known at two loops, both for 
massless~\cite{Aybat:2006wq,Aybat:2006mz} and for massive~\cite{Kidonakis:2009ev,
Ferroglia:2009ep,Ferroglia:2009ii,Mitov:2010xw,Chien:2011wz} particles. In the massless case, 
the problem has an extra symmetry, since the momenta of all hard particles can be 
independently rescaled without affecting infrared emissions. This symmetry puts strong 
constraints on the soft anomalous dimension matrix: a simple solution to these constraints 
is the dipole formula~\cite{Gardi:2009qi,Becher:2009qa}
\beq
  \Gamma_{\rm dip}  \left(\frac{p_i}{\lambda}, \alpha_s(\lambda^2) \right) \, = \,
  \frac{1}{4} \, \widehat{\gamma}_K \left(\alpha_s (\lambda^2) \right) \,
  \sum_{(i,j)} \ln \left(\frac{- s_{i j}}{\lambda^2} 
  \right) {\bf T}_i \cdot {\bf T}_j \, - \, \sum_i
  \gamma_i \left(\alpha_s (\lambda^2) \right) \, ,
\label{sumodipoles}
\eeq
stating that only two-particle color and kinematic correlations contribute to the soft
anomalous dimension. In \eq{sumodipoles} ${\bf T}_i$ is a color generator in the 
representation of particle $i$, and the running coupling enters only through the 
anomalous dimensions $\gamma_i$, responsible for collinear emissions from 
hard parton $i$, and $\widehat{\gamma}_K$, which is the (light-like) cusp 
anomalous dimension~\cite{Korchemsky:1985xj,Korchemsky:1987wg}, rescaled 
by the quadratic Casimir eigenvalue of the relevant color representation.
As a consequence, so long as \eq{sumodipoles} applies, the path ordering operator
${\cal P}$ can be omitted in \eq{RGsol}. The dipole formula is exact at two loops,
and it is known that it can only receive corrections starting with three-loop four-point 
amplitudes, due to the existence at that order of conformally invariant cross-ratios 
of momenta which automatically satisfy the rescaling constraints. Evidence for the 
existence of such corrections, albeit at four loops, was recently uncovered in 
Ref.~\cite{Caron-Huot:2013fea}. Further corrections could arise, starting at four 
loops, if the cusp anomalous dimension $\gamma_K (\alpha_s)$ were to contain 
terms proportional to higher rank Casimir eigenvalues, at that order or beyond.
The nature of the corrections to the dipole formula is currently the subject of intense 
investigations~\cite{Dixon:2009ur,Ahrens:2012qz}.

For massive particles, the simplicity arising from the extra rescaling symmetry is 
lost, and indeed the two-loop calculation shows that tripole correlations among
colored particles do arise. Denoting the Minkowsky angle between hard particles
$i$ and $j$ by
\beq
  \gamma_{i j} \, = \, \frac{2 p_i \cdot p_j}{\sqrt{p_i^2 p_j^2}} \, 
  \equiv \, - \alpha_{i j} - \frac{1}{\alpha_{i j}} \, 
\label{cuspang}
\eeq
and then defining $\xi_{i j} \equiv \log \alpha_{i j}$,  one can express three-particle
correlations in the two-loop soft anomalous dimension as
\beq
  \Gamma_{\rm trip}^{(2)} \left( \xi_{mn} \right) \, = \, {\rm i} f_{a b c} \sum_{i j k} 
  \, {\bf T}_i^a {\bf T}_j^b {\bf T}_k^c \,\, {\cal F}^{(2)} \left( \xi_{ij}, \xi_{jk}, \xi_{ki} \right) \, ,
\label{trip}
\eeq
where~\cite{Ferroglia:2009ep}
\beq
  {\cal F}^{(2)} \left( \xi_{ij}, \xi_{jk}, \xi_{ki} \right) \, = \,  \frac{4}{3} \, \sum_{i j k} 
  \epsilon_{i j k} \,\, g \left(\xi_{ij}\right) \xi_{j k} \coth \left( \xi_{j k} \right) \, .
\label{tripfact}
\eeq
Remarkably, \eq{tripfact} has a factorized form in terms of relatively simple functions
of individual cusp angles: indeed, $g$ is a function of uniform weight $w = 2$, while 
the factor $\xi_{i j} \coth \xi_{i j}$ is proportional to the one-loop non-light-like
cusp anomalous dimension. Clearly, also in the massive case, there is a structural
simplicity which has yet to be fully understood.

\section{On the high-energy limit}
\label{high}

As a first example of application of our understanding of long-distance singularities,
let's consider the high-energy limit. Focusing on four-point amplitudes, this is the limit
in which $s \to \infty$, while $|t|$ remains limited. To the extent that the dipole formula,
\eq{sumodipoles}, applies, the infrared operator ${\cal Z}$ in this limit simplifies considerably,
taking on the form~\cite{DelDuca:2011ae,Bret:2011xm,DelDuca:2013ara,us}
\beq
  {\cal Z} \left(\frac{s}{\mu^2}, \frac{t}{\mu^2}, \as \right) \, = \, 
  {\cal Z}_{1, {\bf R}} \left(\frac{t}{\mu^2}, \as \right) \,   
  \widetilde{\cal Z}_S \left(\frac{s}{t}, \as \right) + \ord \left(\frac{t}{s} \right) \, ,
\label{ZfactS}
\eeq
where $\widetilde{\cal Z}_S$ is a matrix carrying all the leading-power energy dependence,
while ${\cal Z}_{1, {\bf R}}$ is a color-singlet factor. Introducing `Mandelstam' combinations
of color generators, ${\bf T}_s \equiv {\bf T}_1 + {\bf T}_2$, and similarly for ${\bf T}_t$
and ${\bf T}_u$, one can write the operator $\widetilde{\cal Z}_S$ as
\beq
  \widetilde{{\cal Z}}_{\rm S} \left( \frac{s}{t}, \as \right)
  \, = \, \exp \Bigg\{ K( \as ) \bigg[ \bigg( \log \left( \frac{s}{-t } 
  \right) - {\rm i} \, \frac{\pi}{2} (1+ \kappa_{a b}  ) \bigg){\bf T}_t^2
  + {\rm i} \, \frac{\pi}{2} \, \left({\bf T}_s^2 - {\bf T}_u^2 + 
  \kappa_{a b} {\bf T}_t^2  \right) \bigg] \Bigg\} \, .   
\label{Zfact2} 
\eeq
The factor $\kappa_{a b}$ distinguishes quark and gluon amplitudes, with $a,b = q,g$, 
and one has $\kappa_{g g} = \kappa_{q g} = 1$, while $\kappa_{q q} = 4/N_c^2 - 1$, 
due to the different symmetry properties of the color factors of the respective amplitudes. 
$K(\alpha_s)$, on the other hand, is given by a scale integral of the cusp anomalous 
dimension~\cite{Korchemsky:1993hr,Korchemskaya:1994qp,Korchemskaya:1996je,
Magnea:2000ss},
\beq
  K \left( \as \right) = 
  - \frac{1}{4} \int_0^{\mu^2} \frac{d \lambda^2}{\lambda^2} \,
  \hat{\gamma}_K \left( \alpha_s (\lambda^2) \right) \, .
\label{cusp}
\eeq
Finally, the interesting feature of the color-singlet operator ${\cal Z}_{1, {\bf R}}$, whose
explicit expression can be found in~\cite{DelDuca:2013ara}, is that it can be written as 
a product of factors, each associated to one of the external legs. Distinguishing again 
quark and gluon scattering one can write
\beq
  {\cal Z}_{{\bf 1}, {\bf R}}^{a b}  \left( \frac{t}{\mu^2}, \as \right) 
  \, = \,  \left({\cal Z}_{{\bf 1},{\bf R}}^a \left( \frac{t}{\mu^2}, \as \right)\right)^2
           \left({\cal Z}_{{\bf 1},{\bf R}}^b \left( \frac{t}{\mu^2}, \as \right)\right)^2 \, ,
\label{jetfactors}
\eeq
making it natural to think of each external leg contribution as a `jet' factor. These results
of infrared factorization at high energy must be compared to the expressions one gets
from Regge factorization~\cite{Collins:1977jy,DelDuca:1995hf}, under the common assumptions 
that the only singularities of these amplitudes in the complex angular momentum plane should 
be Regge poles. Using these methods one finds that the $t$-channel color octet components 
of quark and gluon amplitudes can be written as~\cite{Balitsky:1979ap,Fadin:1993wh}
\beqa
  {\cal M}_{a b} ^{[8]} \left(\frac{s}{\mu^2}, \frac{t}{\mu^2}, \as \right)
  & = & 2 \pi \alpha_s \, H^{(0),[8]}_{a b}   
  \nonumber \\ && \hspace{-2cm} \times \,\,
  \Bigg\{
  C_a \left(\frac{t}{\mu^2}, \as \right)
  \bigg[ A_+ \left(\frac{s}{t}, \as \right) + \, \kappa_{a b}  \,
  A_- \left(\frac{s}{t}, \as \right) \bigg]
  C_{\rm b} \left(\frac{t}{\mu^2}, \as \right) \nonumber \\
  & & \, + \, \, {\cal R}_{a b}^{[8]} \left(\frac{s}{\mu^2}, \frac{t}{\mu^2}, \as \right)
 + \ord \left( \frac{t}{s} \right) \Bigg\} \, ,
\label{ReggeFact}
\eeqa
where the functions $C_a$ are impact factors, which depend on the identity of
the particles undergoing the high-energy scattering, $H^{(0),[8]}_{a b}$ is the
matrix element for tree-level octet exchange, and the (logarithmic) energy 
dependence is contained in the `Regge trajectory' factor
\beq
  A_\pm \left(\frac{s}{t}, \as \right) \, = \, \left( \frac{- s}{- t} \right)^{\alpha(t)}
  \pm \left( \frac{s}{-t} \right)^{\alpha(t)} \, .
\label{ReggeStructure}
\eeq
We have also included in \eq{ReggeFact} an octet `remainder' function ${\cal R}_{a b}^{[8]}$,
accounting for the fact that \eq{ReggeFact} is not exact at leading power, but only to
leading logarithmic accuracy, and at NLL for the real part of the amplitude~\cite{Fadin:2006bj}. 
As we will see below, infrared factorization allows us to extract useful information on the 
remainder, going beyond the contributions of Regge poles.

\section{Comparing factorizations}
\label{compa}

The simplest way to compare the two high-energy factorizations, \eq{IRfact} and
\eq{ReggeFact}, is to expand the matrix element in powers of the coupling and in 
powers of the high-energy logarithm, as
\beq
  {\cal M}^{[j]} \left(\frac{s}{\mu^2}, \frac{t}{\mu^2}, \as \right) \, = \, 4 \pi \as \, 
  \sum_{n = 0}^\infty \sum_{i = 0}^n
  \left( \frac{\as}{\pi} \right)^n \ln^i \left( \frac{s}{- t} \right)
  M^{(n), i, [j]} \left( \frac{t}{\mu^2} \right) \, ,
\label{AmpExpansion}
\eeq
and perform similar expansion for the various factors on the right-hand sides of
\eq{IRfact} and \eq{ReggeFact}. Thus, for any function involving high-energy logarithms,
$(n)$ will denote the perturbative order, $i$ the power of  $\log(s/|t|)$, and $[j]$ the
color representation in a $t$-channel basis. Even before expanding, however, a direct 
comparison of \eq{IRfact} and \eq{ReggeFact} at leading logarithmic (LL) level shows
that the Regge trajectory $\alpha(t)$ must be strictly related to the cusp integral, \eq{cusp}.
More precisely, one finds that if an amplitude, at tree-level, and in the high-energy 
limit, is dominated by $t$-channel exchange of a particle in representation $[j]$ of 
the gauge group, then that particle will undergo the process of `Reggeization', 
and the divergent contributions to its Regge trajectory will be given 
by~\cite{DelDuca:2011ae,Bret:2011xm,Korchemskaya:1996je}
\beq
  \alpha^{[j]} \left( \frac{t}{\mu}, \alpha_s (\mu) \right) \, = \, 
  {\cal C}^{[j]}_2 \, K \left( \as(-t) \right) \, ,
\label{Reggediv}
\eeq  
where ${\cal C}^{[j]}_2$ is the quadratic Casimir eigenvalue of representation $[j]$. 

A more detailed analysis allows one to begin identifying the terms that break the 
Regge-pole-based factorization \eq{ReggeFact}, starting at two loops and at
NNLL accuracy. Doing this is quite interesting, since at this level a more general
picture of Reggeization is expected to arise, involving Regge cuts in the complex 
angular momentum plane. A precise identification of the contributions to the amplitude 
which violate \eq{ReggeFact} may provide key information to jump-start a new
kind of high-energy resummation. In this spirit, one observes that at two loops 
\eq{IRfact} suggests a natural expression for impact factors~\cite{us},
\beq
  C_a^{(2)} \, = \, \frac{1}{2} Z^{(2)}_{1, {\bf R}, aa} - \frac{1}{8} 
  \left( Z^{(1)}_{1, {\bf R}, aa} \right)^2 + \frac{1}{4} Z^{(1)}_{1, {\bf R}, aa} \, 
  \RE \left[ H^{(1), 0, [8]}_{aa}/H^{(0),[8]}_{a a} \right] \, + {\cal O} \left( \epsilon^0 \right) \, ,
\label{newC}
\eeq
simply arising from the expansion of the `jet factors' in \eq{jetfactors}, acting on the 
finite factors $H^{(n),[8]}$. A direct matching of the two factorizations, however, gives
a much more intricate expression for the impact factor, which evidently breaks
universality, involving mixing with non-octet components and depending on the 
identities of all particles involved in the scattering. The expressions derived from 
IR factorization allow for a precise identification of non-universal terms, which are
naturally assigned to the `remainder' function ${\cal R}^{[8]}$. We can check the 
consistency of our approach by computing the newly defined remainders for 
quark and gluon amplitudes, which yields double poles of the form~\cite{DelDuca:2013ara}
\beq
  R^{(2), 0, [8]}_{qq} \, = \, \frac{\pi^2}{4 \epsilon^2}
  \left(1 - \frac{3}{N_c^2} \right) \, , \quad \, 
  R^{(2), 0, [8]}_{gg} \, = \,  - \, \frac{ 3 \pi^2}{2 \epsilon^2} \, , \quad \,
  R^{(2), 0, [8]}_{qg} \, = \, - \, \frac{\pi^2}{4 \epsilon^2}
  \,. \nonumber
\eeq
Next, we can construct a function measuring the discrepancy between the predictions 
of Regge factorization for the quark-gluon amplitude, which are based on universality, 
and the actual matrix elements. We find~\cite{DelDuca:2013ara}
\beqa
  \Delta_{(2),0,[8]} & \equiv & \frac{M^{(2),0}_{qg}}{H^{(0),[8]}_{qg}} - 
  \bigg[C^{(2)}_q + C^{(2)}_g + C^{(1)}_q C^{(1)}_g - \frac{\pi^2}{4} 
  \left(1 + \kappa_{q g} \right) (\alpha^{(1)})^2 \bigg] \nonumber \\
  & = & \frac{1}{2}\bigg[ R^{(2), 0, [8]}_{qg} - \frac{1}{2} 
  \left( R^{(2), 0, [8]}_{qq} + R^{(2), 0, [8]}_{gg} \right)\bigg] 
  \, = \, \frac{\pi^2}{\eps^2} \frac{3}{16} \left(\frac{N_c^2 + 1}{N_c^2} \right)\, ,
\label{delta}
\eeqa
which precisely reproduces the result of Ref.~\cite{DelDuca:2001gu}, where a failure
of \eq{ReggeFact} was first observed.

Moving on to the three-loops, one finds, as expected, that the breaking of universality
occurs at the level of single-logarithmic terms. Indeed, if one attempts to find an 
expression for the three-loop Regge trajectory using soft-collinear ingredients, one
finds a set of non-universal terms, involving both color mixing and process-dependent 
contributions. Following our general strategy, we assign these terms to the remainder
functions. This gives a prediction for singular, single-logarithmic terms in three-loop 
quark and gluon amplitudes, which break Regge universality. As an example, triple 
pole contributions are given by~\cite{DelDuca:2013ara}
\beq
  R^{(3), 1, [8]}_{qq} \, = \, \frac{\pi^2}{\epsilon^3} \, 
  \frac{2 N_c^2 - 5}{12 N_c} \, , \quad
  R^{(3), 1, [8]}_{gg} \, = \, - \, \frac{\pi^2}{\epsilon^3} \,
  \frac{2}{3} \, N_c \, , \quad
  R^{(3), 1, [8]}_{qg} \, = \, - \, \frac{\pi^2}{\epsilon^3} \,
  \frac{N_c}{24} \, .
\label{explR3}
\eeq
Single pole contributions can be computed as well, but they are, as might be expected,
considerably lengthier, involving interference with finite two-loop contributions as well. 

We conclude this section by noting that, just as infrared factorization provides important
information in the high-energy limit, one may also use Regge factorization to extract
constraints on the finite parts of the amplitudes, which are not in principle controlled by 
\eq{IRfact}. Specifically, and interestingly, one may show~\cite{us} that LL and
NLL octet hard parts of quark and gluon amplitudes vanish in dimensional regularization
to all orders in perturbation theory. More precisely one finds that
\beqa
  {\rm Im} \left( H^{(n), n, [8]} \right) & = & 0 \, , \qquad \qquad \quad \!
  {\rm Re} \left( H^{(n), n, [8]} \right) \, = \, O(\epsilon^n) \, , \label{allord} \\
  {\rm Im} \left( H^{(n), n - 1, [8]} \right) & = & O(\epsilon^n) \, , \qquad
  {\rm Re} \left( H^{(n), n - 1, [8]} \right) \, = \, {\cal O} (\epsilon ^{n-2}) \, ,
  \nonumber 
\eeqa
where the only finite order information used is the vanishing of the one-loop octet
hard part. \eq{allord} reinforces the idea that high-energy logarithms are infrared in 
nature: thus, infrared-finite high-energy logarithms must come from the interference 
of soft and collinear functions with lower-order contributions subleading in $\epsilon$. 
These constraints, discussed in greater detail in~\cite{us}, are the subject of ongoing 
investigations.

\section{Weaving webs}
\label{webs}

Another direction of investigation which has seen remarkable progress in recent 
years is the development of techniques for the direct calculation of the soft anomalous
dimension. The key element of these techniques is the fact that soft gluon contributions
to the infrared operator ${\cal Z}$ can be expressed in terms of correlators of Wilson
lines, and such correlators are known to exponentiate. Furthermore, diagrammatic 
rules exist that allow one to compute directly the logarithm of the correlator, considerably 
simplifying the task of extracting the anomalous dimension. For abelian gauge theories,
the exponent is very simple, involving only connected diagrams~\cite{Yennie:1961ad}, 
and thus only correlations induced by matter loops. For non-abelian theories, in the case 
of amplitudes involving only two hard partons, this rule generalizes~\cite{Sterman:1981jc,
Gatheral:1983cz,Frenkel:1984pz} in a relatively simple way: the exponent is built out of 
diagrams, called `webs', which are two-Wilson-line-irreducible, {\it i.e.} that cannot be 
partitioned into disconnected subdiagrams by cutting only the two Wilson lines. 
These diagrams then enter the exponent with modified color factors which can be 
recursively computed from the ordinary Feynman rules. The general case of multi-particle 
non-abelian amplitudes is considerably more intricate~\cite{Laenen:2008gt,Gardi:2010rn,
Gardi:2011wa,Gardi:2011yz,Gardi:2013ita,Mitov:2010rp}. In particular, the concept of 
`web' must be rather drastically generalized: webs are in this case sets of Feynman 
diagrams, which are closed under the operation of permuting the order of their attachments 
to each Wilson line. Writing each diagram $D$ as the product of its color factor $C(D)$ 
and its kinematic integral ${\cal F} (D)$, one finds that each web contributes to the 
logarithm of the correlator through the combination
\beq 
  W \, = \, \sum_{D,D' \in W} {\cal F}(D) R_{DD'} C(D') \, ,
\label{Wdef}
\eeq
where the sum runs over all diagrams $D$, $D'$ contributing to the web $W$, and 
$R_{D D'}$ is a `web mixing matrix' whose entries are rational numbers of combinatorial 
origin.

The physically interesting information in the Wilson line correlator is contained in its
ultraviolet divergences, which can be mapped to the infrared divergences of the original
scattering amplitude. Extracting these divergences is non-trivial, due to the highly
singular nature of these correlators, which contain their own infrared divergences, 
as well as collinear divergences in the massless case. To disentagle singularities of 
different physical origin, it is useful to consider non-lightlike Wilson lines (which are 
of course also directly related to infrared singularities of massive colored particles),
and introduce a mass scale $m$ as an infrared regulator. Dimensional regularization
then controls the UV singularities which are the target of the calculation. One considers
therefore the regulated correlator
\beq
  {\cal S} \left( \gamma_{ij}, \alpha_s(\mu), \e, \frac{m}{\mu} \right) \, \equiv \, 
  \left\langle 0 \left| \Phi_{\beta_1}^{(m)} \otimes \Phi_{\beta_2}^{(m)} \otimes
  \ldots \otimes \Phi_{\beta_L}^{(m)} \right| 0 \right\rangle \, ,
\label{Sdef}
\eeq
where $\gamma_{i j}$ are the cusp angles defined in \eq{cuspang}, and each Wilson line 
$\Phi_{\beta_i}$, pointing in the direction $\beta_i$ defined by the momentum $p_i$ of the 
$i$-th particle, is regulated at large distances by an exponential regulator~\cite{Gardi:2013saa}, 
as
\beq
  \Phi_{\beta_i}^{(m)} \, = \, {\cal P} \exp \left[ {\rm i} g \mu^\e 
  \int_0^\infty d \lambda \, \beta_i \cdot A \left( \lambda \beta_i \right) \, 
  {\rm e}^{- m \lambda \sqrt{\beta_i^2}}
  \right] \, ,
\label{Phidef}
\eeq
where we take momenta to be time-like, $\beta_i^2 > 0$. The multiplicative renormalizability
of Wilson line correlators now ensures that a UV finite version of ${\cal S}$ can be 
constructed,
\beq
  {\cal S}_{\rm ren.} \left( \gamma_{ij}, \alpha_s(\mu^2), \e, \frac{m}{\mu} 
  \right) \, = \, {\cal S} \left( \gamma_{ij}, \alpha_s(\mu^2), \e, \frac{m}{\mu} 
  \right) \, Z \left( \gamma_{ij}, \alpha_s(\mu^2), \e \right) \, ,
\label{Srendef}
\eeq
with the matrix renormalization factor $Z$ containing all physically relevant information.
The matrix nature of $Z$ entails a further subtlety: in order to properly subtract the UV
divergences of ${\cal S}$, the expression for $Z$ at high orders must include commutators
of lower-order contributions, as dictated by the Baker-Campbell-Haussdorf formula. The 
structure of these commutator counterterms is easily worked out: for example, up to three 
loops, one finds that the renormalization factor $Z$ can be expressed in terms of the soft
anomalous dimension $\Gamma$ as
\beqa
  Z \left(\alpha_s, \e \right) & = & \exp \Bigg\{ \alpha_s \,
  \frac{1}{2 \e} \, \Gamma^{(1)} + \, \alpha_s^2 
  \left( \frac{1}{4 \e} \Gamma^{(2)} - \frac{b_0}{4 \e^2} \Gamma^{(1)} \right) \\ 
  & & + \, \alpha_s^3 \left( \frac{1}{6 \e} \Gamma^{(3)} + 
  \frac{1}{48 \e^2} \left[ \Gamma^{(1)}, \Gamma^{(2)} \right] - 
  \frac{1}{6 \e^2} \left(b_0 \Gamma^{(2)} + b_1 \Gamma^{(1)} \right) + 
  \frac{b_0^2}{6 \e^3} \Gamma^{(1)} \right) \Bigg\} \, ,
\label{Zexp}
\eeqa
where $\Gamma^{(n)}$ is the $n$-th order perturbative coefficient of $\Gamma (\alpha_s)$.
One sees that in the non-abelian case the logarithm of $Z$ involves multiple UV poles,
even in the conformal limit in which the coefficients $b_n$ of the $\beta$ function vanish.

Armed with these tools, computing $\Gamma$ is a well defined, if lengthy and rather 
intricate, task. It is sufficient to compute the regularized soft function ${\cal S}$, which 
can be done directly at the level of the exponent, as
\beq
  {\cal S} \left(\alpha_s, \e \right) \, = \, \exp \Big[ w \left( \alpha_s, \e \right) \Big]
  \, = \, \exp \left[ \, \sum_{n = 1}^\infty \, \sum_{k = - n}^\infty  \alpha_s^n \, \e^k \, 
  w^{(n, k)} \right] \, .
\label{Sunren}
\eeq
The anomalous dimension is now constructed in terms of combinations of the perturbative
coefficients $w^{(n, k)}$, including commutator counterterms. Up to three loops, one gets
\beqa
\label{Gamres}
  \Gamma^{(1)} & = & - 2 w^{(1,-1)} \, ,\nonumber \\
  \Gamma^{(2)} & = & - 4 w^{(2,-1)} - 2 \left[ w^{(1,-1)}, w^{(1,0)} \right] \, ,\nonumber \\
  \Gamma^{(3)} & = & - 6 w^{(3,-1)} + \frac{3}{2} b_0 \left[ w^{(1,-1)}, w^{(1,1)} \right]
  + 3 \left[ w^{(1,0)}, w^{(2,-1)} \right] + 3 \left[ w^{(2,0)}, w^{(1,-1)} \right] \nonumber \\
  & & \hspace{-1cm} 
  + \left[ w^{(1,0)}, \left[w^{(1,-1)}, w^{(1,0)} \right] \right] - \left[ w^{(1,-1)}, 
  \left[w^{(1,-1)}, w^{(1,1)} \right] \right] \, .
\label{Gamma}
\eeqa
Notice that, while $\Gamma^{(n)}$ is given, as expected, by a combination of single pole
and $\beta$ function contributions, also positive powers of $\epsilon$ in the logarithm 
of the regularized correlator play a role. Notice also that the matrix $w ( \alpha_s, \e)$
has a non-trivial structure: order by order, it is the sum of all relevant webs, each of 
which in general contributes to several color structures. Furthermore, most webs have
multiple UV poles, and therefore are not directly physical: they depend on the IR cutoff
$m$. Only when appropriate commutator subtractions have been included one is left with
a physical contribution to $\Gamma$, corresponding to a single UV pole, and thus 
independent of $m$.

\section{Multiple gluon exchange webs}
\label{MGEWs}

I would like to conclude this contribution by giving a few details about a class of webs
which has recently received a lot of attention~\cite{Gardi:2013saa,Falcioni:2014pka,
Henn:2013wfa}. These are the webs which are generated if one computes the soft 
function ${\cal S}$ with a path integral weighted by just the quadratic terms in the action
\beq
  \left. {\cal S} \left( \gamma_{ij}, \alpha_s(\mu), \e, \frac{m}{\mu} \right) 
  \right|_{\rm MGEW} \, \equiv \, \int \left[ D A \right] \Phi_{\beta_1}^{(m)} 
  \otimes \Phi_{\beta_2}^{(m)} \otimes \ldots \otimes \Phi_{\beta_L}^{(m)}
  \, \exp \Big\{ {\rm i} S_0 [A] \Big\} \, .
\label{formdef}
\eeq
In diagrammatic terms, these webs involve graphs in which all gluons attach directly
to the Wilson lines, and there are no three- or four-gluon vertices: they are called
`Multiple Gluon Exchange Webs' (MGEWs). While their graphical appearance is abelian,
it is important to realize that, due to importance of the ordering of gluon insertions on 
the Wilson lines, they have a true non-abelian character, and they contribute to the 
same color structures as fully connected webs involving gluon self-interactions.

\begin{figure}[htb]
\begin{center}
\scalebox{0.8}{\includegraphics{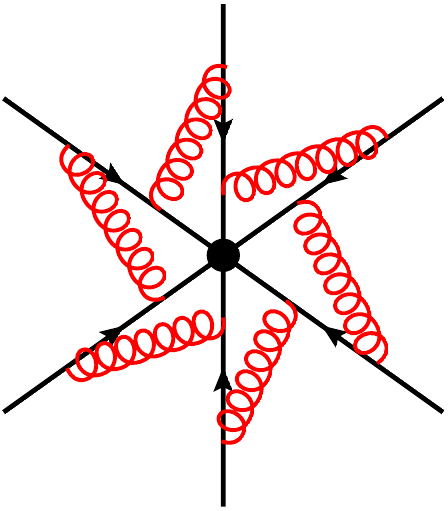}}
\caption{Example of an Escher staircase with six external legs.}
\label{EscherFig}
\end{center}
\end{figure}

MGEWs are sufficiently well understood that several of their properties are known or 
conjectured to all orders in perturbation theory, and in fact, as we will mention below,
certain diagrams in this class are explicitly known for any number of gluons. To begin 
with, kinematic factors for diagrams contributing to MGEWs admit a universal 
parametrization, which leads to an explicit, completely general integral representation.
One can write, for any diagram $D$ contributing to a MGEW at $n$ loops,
\beq
  {\cal F}^{(n)} \left( D \right) \, = \, \kappa^n \, \, \Gamma( 2 n \e ) 
  \, \int_0^1 \prod_{k = 1}^n \Big[ d x_k  \,\gamma_k \, P_\epsilon 
  \left( x_k, \gamma_k \right) \Big]  \, \phi_D^{(n)} \left( x_i; \e \right)  
  \nonumber \, .
\label{gendiag3} 
\eeq
Here $\kappa = - g^2/(8 \pi^2) + {\cal O} (\epsilon)$, $\gamma_k$ is the cusp angle
subtended by the $k$-th gluon, and $x_k$ is an angle-like variable measuring the
degree of collinearity of the $k$-th gluon with the emitting and the absorbing Wilson 
lines. $P_\epsilon (x_k, \gamma_k)$ is a function arising from the coordinate space
gluon propagator, and given by
\beq
  P_\epsilon \left( x, \gamma \right) \, \equiv \, 
 \left[ x^2  +(1 - x)^2 -  x (1 - x)
 \gamma) \right]^{-1 + \e} \, ;
\label{propafu}
\eeq
finally, the kernel $\phi_D^{(n)} (x_i; \e)$ contains the information about the ordering 
of gluon attachments on each Wilson line, which is specific to diagram $D$; one may 
write
\beq
  \phi_D^{(n)} \left( x_i; \e \right) \, = \, \int_0^1 \prod_{k = 1}^{n - 1} d y_k  
  \left( 1 - y_k \right)^{- 1 + 2 \e} \, y_k ^{- 1+ 2 k \e} \, \, 
  \Theta_D \big[ \left\{ x_k, y_k \right\} \big] \, ,
\label{phi_D}
\eeq
where $\Theta_D [ \{ x_k, y_k \} ]$ is a product of Heaviside functions enforcing the
proper ordering on each line. The integral representation in \eq{gendiag3} is sufficiently
general and manageable to allow for calculations at very high orders. To substantiate this,
consider the special class of highly symmetric $n$-loop MGE diagrams illustrated in 
Fig.~\ref{EscherFig} for $n = 6$, which were called `Escher staircases' in \cite{Gardi:2010rn}.
At each order, there are two such diagrams, enantiomers of each other. As is apparent 
from their graphical structure, these diagrams have no subdivergences, and thus
yield directly a single pole, contributing to the anomalous dimension, without the need 
for subtractions. Remarkably, these diagrams can be computed for any $n$, and give
a very simple result. To display it, define
\beq
  S_n (x_i) \, \equiv \, \prod_{i = 1}^n \frac{1 - x_i}{x_i} \, .
\label{Sn}
\eeq
as well as
\beq
  \theta_+ (n) \, \equiv \, \theta \Big( S_n (x_i) - 1 \Big) \, ,
\label{theta+}
\eeq
and $\theta_- (n) \equiv 1 - \theta_+ (n)$, which define the overall orientation 
of the staircase. In terms of these simple ingredients, one finds that the kernel 
(\ref{phi_D}) of the staircase diagrams for any $n$ can be written in the remarkably 
compact form
\beq
  \phi_S^{(n)} \left( x_i; \, 0 \right) \, = \, \frac{1}{(n - 1)!} \, \, \theta_+ (n)  \, 
  \bigg( \log \Big[ S_n (x_i) \Big] \bigg)^{n - 1} \, .
\label{finstair}
\eeq 
This form was used in Ref.~\cite{Falcioni:2014pka} to prove an all-order result for a physical 
contribution to the anomalous dimension: one finds that the coefficient of a specific color 
structure appearing in the MGEWs which feature the staircase diagrams must vanish for 
any $n$.

In general, beyond simple and symmetric cases like the staircase diagrams, one must
realize that computing the integrals in \eq{gendiag3} is only the first step in a more 
articulate procedure, and several further steps are needed in order to arrive from individual 
diagrams to the anomalous dimension: diagrams must be combined into webs, which 
in general comprise several color structures, and, importantly, commutator counterterms 
must be subtracted. Only at that point multiple poles will cancel, and one can construct
contributions to the physical anomalous dimension. Starting from \eq{gendiag3}, 
and building up the contribution to the anomalous dimension corresponding to 
web $W$ and color structure $j$, after subtraction of counterterms, one finds an 
$\epsilon$-independent expression of the form
\beqa
  F^{(n)}_{W, \, j} \big( \alpha_i \big) & = &
  \int_0^1 \left[ \, \prod_{k = 1}^n d x_k \, p_0 (x_k, \alpha_k) \right]  \, 
  {\cal G}^{(n)}_{W, \, j} \Big(x_i, q(x_i, \alpha_i) \Big) \, ,
\label{subtracted_web_mge_kin}
\eeqa
where the function $p_0$ is the limit of \eq{propafu} as $\epsilon \to 0$, now expressed 
in terms of the variable $\alpha$ defined in \eq{cuspang}, while $q(x, \alpha)$ is the 
logarithm of the quadratic form in $x$ appearing in \eq{propafu}. This logarithm arises 
when \eq{propafu} is expanded in powers of $\epsilon$ and combined with higher-order 
poles arising in individual diagrams. At this level, and at this level only, the structure
of the answer simplifies drastically, and its analytic properties become apparent. One 
finds, for all webs computed so far, up to four loops and five external legs, that the
function $F^{n}_{W, \, j}$ obeys the following key properties, which were conjectured 
in~\cite{Gardi:2013saa} to be valid for all MGEWs: first, $F^{n}$ is {\it factorized}, that is, it is 
a sum of terms, each of weight $2 n - 1$, and each given by a product of functions 
depending on individual cusp angles; second, the {\it symbol} of $F^{n}$ is built out of a
very simple alphabet, comprising only the letters $\{ \alpha_i, 1 - \alpha_i^2 \}$, with 
$i = 1, \ldots, n$.

These general insights, paired with high-order and all-order examples, suggest that
the problem of MGEWs may be solvable in closed form. In parallel, efforts are 
continuing to compute other, more connected webs contributing to the three-loop 
soft anomalous dimension for both massless and massive particle scattering. The
remarkable progress that has been achieved in the past few years makes it likely
that this problem too will be completely solved in the not too far future, a statement 
that would have been considered quite implausible in previous iterations of
this Conference.

\vspace{0.5cm}

{\large{\bf Acknowledgements}}

\vspace{2mm}

\noindent Work supported by MIUR (Italy), under contract 2010YJ2NYW$\_$006 and 
by the University of Torino and the Compagnia di San Paolo under contract ORTO11TPXK.
The author thanks Vittorio Del Duca, Giulio Falcioni, Einan Gardi, Mark Harley, Leonardo 
Vernazza and  Chris White for the fruitful collaborations that yielded some of the results 
reviewed here.

\end{document}